\documentclass[epj]{svjour}

\usepackage{graphics}
\usepackage[dvipsnames]{xcolor}	
\usepackage{bm}

\newcommand{\tobs}{t_{\rm obs}}
\newcommand{\mtobs}{\mu \tau_{\rm obs}}
\newcommand{\mstar}{\mu^{\ast} \tau_{\rm obs}}
\newcommand{\Ts}{T_{\rm s}}
\newcommand{\etax}[1]{\bm{\eta}_{#1}}
\newcommand{\kex}{\kappa_{\rm exc}}
\newcommand{\ft}[1]{{\color{black} #1}}

\begin{document}\sloppy
\title{Structural-dynamical transition in the Wahnstr\"om mixture}

\author{Francesco Turci\inst{1,2} \and Thomas Speck \inst{3} \and C. Patrick Royall\inst{1,2,4}}

\institute{H.H. Wills Physics Laboratory, University of Bristol, Bristol, BS8 1TL, United Kingdom \and Centre for Nanoscience and Quantum Information, Bristol BS8 1FD,
  United Kingdom  \and Institut f\"ur Physik, Johannes Gutenberg-Universit\"at Mainz,  Staudingerweg 7-9, 55128 Mainz, Germany \and School of Chemistry, University of Bristol, Bristol BS8 1TS, United Kingdom }

%

%
\date{Received: date / Revised version: date}
%
\abstract{
In trajectory space, dynamical heterogeneities in glass-forming liquids correspond to the emergence of a dynamical phase transition between an \ft{active} phase poor in local structure and an \ft{inactive} phase which is rich in local structure. We support this scenario with the study of a model additive mixture of Lennard-Jones particles, quantifying how the choice of the relevant structural and dynamical observable affects the transition in trajectory space. We find that the low mobility, structure-rich phase is dominated by icosahedral order.  Applying a nonequilibrium rheological protocol, we connect local order to the emergence of mechanical rigidity.
\PACS{
      {PACS-64.70.Q-}{theory and modelling of glasses}   \and
      {PACS-83.10.Rs}{computer simulation of molecular dynamics in rheology}
     } 
} 
\maketitle

\section{Introduction}
\label{sec:intro}
Supercooled liquids show emergent dynamical and structural heterogeneities when cooled towards the glass transition \cite{cavagna2009,berthier2011,royall2015physrep,barrat2011}. The relation between slow dynamics and some form of short-range (local) order, however, is still poorly understood. On the one hand, the efficient filling of space with atoms of different sizes requires a certain degree of topological order \cite{torquato2000} and the dynamic slowdown can rigorously be linked to emerging static lengthscales \cite{montanari2006}; on the other hand, computer simulations have shown that the correlation between local structural features and slow dynamics is strongly model dependent \cite{hocky2014pre,royall2015}. In experiments, colloidal \cite{royall2008,wochner2009,leocmach2012,royall2018jcp} and metallic glasses \cite{sheng2006,chen2011,hirata2013} provide evidence for emerging local order as well as, on the contrary, support for purely dynamical scenarios where local structure has limited influence on the dynamics \cite{gokhale2014,chikkadi2014}. \ft{Historically, the study of local structure with complex higher order metrics has played a decisive role in understanding amorphous systems and packings since the times of Bernal and Finney \cite{bernal1959,bernal1960,finney1970,finney1970mc} and has contributed to a geometric and thermodynamic interpretation of the emerging frustration in glasses \cite{nelson1983,tarjus2005}. However, alternative approaches which disregard structural features and focus on dynamical \cite{chandler2010} or vibrational/elastic aspects \cite{dyre2006} of the problem or relaxation have been proposed, in striking contrast with established thermodynamic theories of the glass transition  \cite{lubchenko2007,lubchenko2015theory}}. It is therefore important to understand what drives strong or weak coupling between structure and dynamics in different supercooled liquids. 

A major difficulty encountered in the investigation of the role of structural changes in dynamic arrest is the fact that particle-resolved studies (and in particular conventional computer simulations) can only access a limited dynamic range of slow relaxation. Typically, this encompasses 4 to 5 orders of magnitude in time, meaning that such studies mainly capture the onset of the mechanisms that characterise the deeply supercooled and glassy regimes (when the relaxation times are 10 to 20 orders of magnitude larger with respect to the liquid regime) \cite{berthier2016}. Therefore, alternative sampling routes to explore the deeply supercooled regime from a structural and/or dynamical point of view have been developed in recent years, including pinning fields \cite{cammarota2012,kob2013,kob2014nonlinear}, particle-swap Monte-Carlo on particular models \cite{berthier2016prl,ninarello2017} or biased dynamical ensembles \cite{chandler2010,hedges2009,speck2012,turci2017prx}. 

A potential route to study dynamical and structural heterogeneities in glassformers is provided by efficient sampling methods in trajectory space, where novel dynamical phase transitions have been uncovered and connected to the dynamical slowdown observed in supercooled liquids \cite{chandler2010}. The study of trajectory space in glassy systems has been originally promoted in the context of the dynamical facilitation theory of slow dynamics \cite{garrahan2007,keys2011,hedges2009}. Within this framework, on-lattice idealised models \cite{garrahan2007,elmatad2009,turci2011,nemoto2017} as well as more realistic models of structural glasses \cite{hedges2009,speck2012,jack2011,pitard2011dynamic,isobe2016} have been shown to undergo a first-order dynamical phase transition in trajectory space between an \textit{active} phase with high mobility (fast relaxation) and an \textit{inactive} phase with low mobility (slow dynamics).

However, this purely dynamical picture has been more recently complemented by a structural aspect: active/inactive phases correspond to trajectories particularly poor/rich in local structure \cite{speck2012jcp,pinchaipat2017} and can be seen as representative of the low temperature state of the supercooled liquid \cite{turci2017prx}. Dynamical transitions are therefore understood to correspond to a \textit{structural-dynamical} transitions, where the slowdown of the dynamics becomes intimately related to the growth of short-range-order domains.

Still, much of the evidence for structural-dynamical phase transitions in atomistic models of glassformers up to now is restricted to only two model systems (the Kob-Andersen mixture \cite{hedges2009,jack2011,speck2012jcp,speck2012,turci2017prx}, a popular Lennard-Jones mixture with weak structural dynamical correlations \cite{hocky2014}, and the moderately polydisperse hard spheres \cite{pinchaipat2017}). In order to understand how system-dependent this picture is, it is important to extend the scope of these studies to other model systems.

In the present numerical work, we consider the case of a popular atomistic glassformer originally introduced by G\"oran Wahnstr\"om as a simple model for supercooled liquids \cite{wahnstrom1991}. It consists in a binary mixture of Lennard-Jones particles whose parametrization has been found to provide a good model of fragile glasses, with a particularly strong coupling between its slow dynamics and the emergence of local geometrical motifs \cite{hocky2014,malins2013jcp,pinney2015}. These are typically icosahedra, a very common arrangement in simple models of glass-forming liquids composed of spherically symmetric particles.

The article is structured as follows: in Section \ref{sec:methods} we present the model studied and the importance sampling technique employed for trajectory sampling; in Section
\ref{sec:structdyn} we introduce the relevant observables and the phase transitions in trajectory space that can be probed through the dynamical $s$-ensemble and the structural-dynamical $\mu$-ensemble; in Section \ref{sec:shear} we show that it is possible to connect the structural-dynamical transition to the emergence of rigidity in the glass, as the icosahedra-rich phase presents distinctive rheological properties; finally, we conclude the article with an overview of the results and their implications.

%

\section{Model and sampling technique}
\label{sec:methods}

\subsection{The Wahnstr\"om binary mixture}
We study the Wahnstr\"om binary  mixture of Lennard-Jones particles. The model is a 50:50 mixture of large (A) and small (B) particles with parameters $\sigma_{A}/\sigma_{B}=1.2$, $m_A/m_B=2$,  $\epsilon_{A}/\epsilon_B=1$ and cutoff $r_{\rm cut}=2.5\sigma$ at number density $\rho=1.296$. Lengths, temperature and times are reported in units of $\sigma_A$, $\epsilon_A /k_B$ , and $(m_A \sigma_A^2 /\epsilon_A )^{1/2}$, respectively. The mixing rule for the interaction is additive, i.e. it follows the Lorentz-Berthelot rules

\begin{equation}
	\sigma_{AB}=\frac{\sigma_A+\sigma_B}{2},\quad\epsilon_{AB}=\sqrt{{\epsilon_A\epsilon_B}}.
\end{equation}

\begin{figure}[t]
  \centering
  \includegraphics{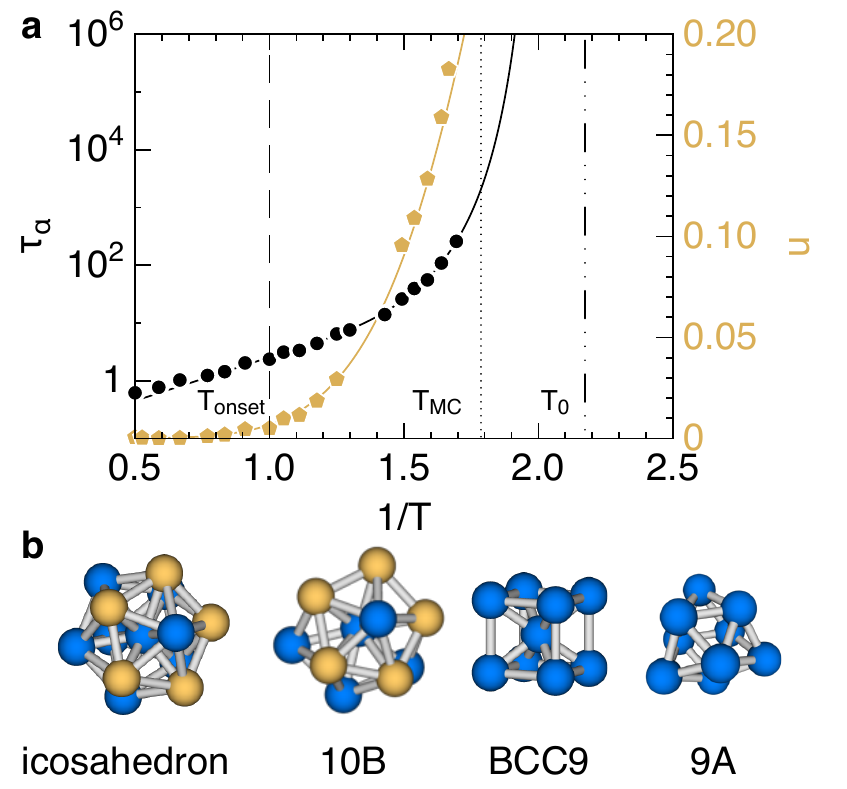}
  \caption{ (a) Structural relaxation time $\tau_{\alpha}$(circles, logarithmic scale) and population of particles in icosahedral motifs $n$ (pentagons, linear scale) as a function of the inverse temperature. Continuous lines are the Vogel-Fulcher-Tammann (VFT) fit $\ln \tau_{\alpha}/\tau_{\infty}=DT_0/(T-T_0)$ and an empirical $n= [1+(T/T_{1/2})^{\gamma}]^{-1}$ fit for $\tau_{\alpha}$ and $n$ respectively, with $\gamma=6.6, T_{1/2}=0.47$ from Ref.~\cite{jenkinson2017}. Vertical lines also indicate the location of relevant temperatures in the Wahnstr\"om model: the onset of slow dynamics $T_{\rm onset}$, the mode-coupling transition temperature $T_{\rm MC}$ and the VFT temperature $T_0$. (b) Three-dimensional rendering of the four local motifs considered in this work: the icosahedron, the defective icosahedron (10B) and two nine-particle motifs unrelated to five-fold symmetry. For the icosahedron and 10B, a pentagonal ring of particles is highlighted in gold.}
  \label{figtauIcovsT}
\end{figure}

This atomistic supercooled liquid has been extensively studied since its original design \cite{wahnstrom1991}. The model reproduces to a good degree the relaxation behaviour of so-called \textit{fragile} glasses, as its structural relaxation time $\tau_{\alpha}$ (as measured from the decay of the intermediate scattering function \cite{malins2013jcp}) undergoes a non-Arrhenius (super-exponential) increase when the system is cooled below the crossover or \textit{onset} temperature $T_{\rm onset}=1.0$ \cite{coslovich2011,malins2013jcp}. Furthermore, as the temperature is decreased, the disordered structure of the liquid changes with the formation of five-fold symmetric domains and in particular of local particle motifs with icosahedral coordination \cite{malins2013jcp,coslovich2011,coslovich2007} which contribute to the emergence of strong frustration \cite{turci2017prl}. Equilibration of the liquid in conventional simulations around and below the so-called mode-coupling temperature $T_{MC}=0.56$ is computationally expensive, making the low temperature, activated regime (crucial for testing theoretical predictions\cite{berthier2011}) unreachable. Divergence of the relaxation times, if modelled by the super-Arrhenius Vogel-Fulcher-Tamman law $\ln \tau_{\alpha}/\tau_{\infty}=DT_0/(T-T_0)$, is predicted at temperature $T_0\approx0.46$. For reference, we report in Fig.~\ref{figtauIcovsT}(a) the temperature dependence of both the structural and dynamical properties of the model.
 
Beyond local structural order, the model has been shown to crystallise, under suitable conditions, into a $\rm MgZn_2$ Laves phase formed by icosahedral motifs and so-called Frank-Kasper bonds \cite{pedersen2010} but in the supercooled liquid regime the contribution of such a large unit cell to the increased degree of local order has been shown to be limited \cite{malins2013jcp}.

\subsection{Replica exchange in trajectory space}

As in previous work \cite{speck2012,turci2017prx,speck2012jcp,pinchaipat2017}, in order to sample large fluctuations of the time-integrated observables we employ an importance sampling technique that extends equilibrium replica exchange methods to ensembles of trajectories.

We sample space and time extensive observables $O_{x}$ on systems of $N=512$ evolving for a finite observation time $\tobs$. A generic time-integrated observable $O_x$ is defined as a double sum over the number of particles and a discretization of time into $L$ intervals for a total of $K=L+1$ points:
\begin{equation}
	O_x = \sum_{t=-L/2}^{+L/2}\sum_{i=1}^N f_{t,i}
\end{equation}
where $f_{t,i}$ is a specific microscopic observable (e.g. a single particle indicator function).

The goal of the importance sampling technique is to efficiently measure the probability distribution $P(O_x;\Ts)$ for a given value of the thermostat temperature $\Ts$. In particular, we are interested in the \textit{large deviations} from the typical value of the probability distribution. In order to calculate such rare fluctuations in trajectory space, new trajectories are generated through shifting and shooting moves (inspired by Transition Path Sampling \cite{bolhuis2002}). Hence, the algorithm performs a random walk in trajectory space with acceptance probability determined by a Metropolis rule
\begin{equation}
\label{eq:acc}
	p_{\rm acceptance} = \min \left\{1, \frac{e^{-\Psi(O_x^{\rm new}}) }{e^{-\Psi(O_x^{\rm old}) }} \right\},
\end{equation}
which ensures detailed balance and where $\Psi(O_x^{\rm new}),\Psi(O_x^{\rm old}) $ are the values of a biasing pseudo-potential which is a function of the extensive observable $O_x$ computed over old and new trajectories. We choose $\Psi$ to have a parabolic form
\begin{equation}
	\Psi_j ( O_x) =\frac{1}{2}\omega (O_x-O_0^j)^2,
\end{equation}
where $O_0^j$ is the reference (typical) value associated to replica $j$. Depending on the observable, we take a number of distinct replicas varying from 8 to 16, with equally spaced values of $O_j$ and values for the harmonic constant $\omega$ that ensure good mixing of neighboring replicas. Mixing is also enhanced by 2500 swap attempts among all (not-necessarily neighboring) replicas.

The Monte-Carlo algorithm in trajectory space simulation starts with an equilibrated trajectory assigned to all replicas at temperature $\Ts$. A new trajectory is then generated via Transition Path Sampling moves (1/4 shifting, 3/4 shooting \cite{bolhuis2002}) independently for every replica, accepted or rejected according to Eq.~\ref{eq:acc}. Swap attempts between different replicas are then performed, completing the cycle. During the sampling, we employ a velocity-Verlet integrator with timestep $dt=0.005$ to resolve the equation of motion and the Andersen thermostat to keep the temperature constant.

\begin{figure}[t]
  \centering
  \includegraphics{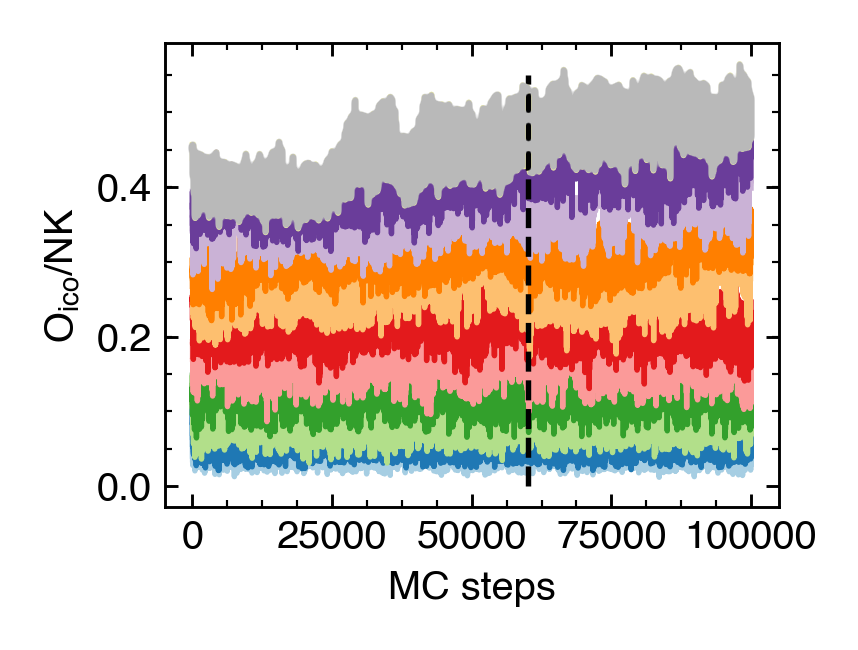}
  \caption{Example of a Monte-Carlo run in trajectory space. At temperature $\Ts=0.71$ we collect trajectories for the time-integrated number of icosahedra from 12 distinct replicas corresponding to the different colours. A very long transient of $6\cdot 10^4$ Monte-Carlo cycles (vertical dashed line) is observed and statistically relevant data are only collected from the remaining $4\cdot10^4$ trajectories.}
  \label{figmc}
\end{figure}

We perform several tens of thousands of cycles and collect statistics and block averages from three to eight non-overlapping blocks of data whose size ranges between $1.2\cdot 10^4$ and $3 \cdot10^4$ trajectories, which deliver estimates for the averages and standard errors. Crucially, depending on the sampling temperature $\Ts$, correlations may be very long-lived and the number of Monte-Carlo cycles spent during the equilibration in trajectory space can be very large ($\sim 6\cdot 10^4$) Monte-Carlo sweeps), as shown in Fig.~\ref{figmc}. We then discard trajectories produced during equilibration and collect data from the converged, late steps of the Monte-Carlo.

From the collected ensemble of trajectories, we calculate distributions and expectation values using the Multistate Bennett Acceptance ratio (MBAR) method extended to ensembles of trajectories \cite{minh2009}. This technique allows us to obtain the unbiased probability distribution $P(O_x;\Ts)$ and expectation values for any quantity $A$ as 

\begin{equation}
\label{eq:averages}
	\langle A\rangle_{y}=\frac{\langle A e^{y O_{x}}\rangle}{\langle e^{y O_x}\rangle}
\end{equation} 
where $y$ is the conjugated field to the observable $O_x$ and $\langle\rangle$ indicate averages  according to the unbiased distribution $P(O_x;\Ts)$. Notice that the denominator in Eq.~\ref{eq:averages} corresponds to the moment generating function of the probability distribution and is a generalization of the partition sum to trajectory spaces. In this work, we focus on two particular ensembles: the $s$-ensemble, where $y=-s$ and the observable $O_x$ is the time-integrated mobility of the particles; and the $\mu$-ensemble, where $y=\mu$ and the relevant observable is a time-integral over the number of particles in a particular local motif (here the icosahedron).

In the presence of transitions in trajectory space, we expect to measure probability distributions for the time-integrated observables that are not Gaussian and display long, eventually exponential tails. For variables that follow a Gaussian probability distribution, the kurtosis $\kappa_4=\langle (O_x-\langle O_x\rangle)^4\rangle/\langle (O_x-\langle O_x\rangle)^2 \rangle^2$ (i.e. the ratio between the fourth central moment and the squared second moment) has value $\kappa_4 = 3$. Therefore, the excess kurtosis $\kappa_{\rm exc}=\kappa_4-3$ is often employed as a benchmark for the deviations from a Gaussian distribution. So-called \textit{leptokurtic} (fat-tailed) distributions correspond to positive $\kex$ while \textit{platykurtic} (thin-tailed) distributions correspond to negative $\kex$.

\section{Dynamical and structural phase transitions}
\label{sec:structdyn}

\begin{figure*}[t]
\centering
  \includegraphics{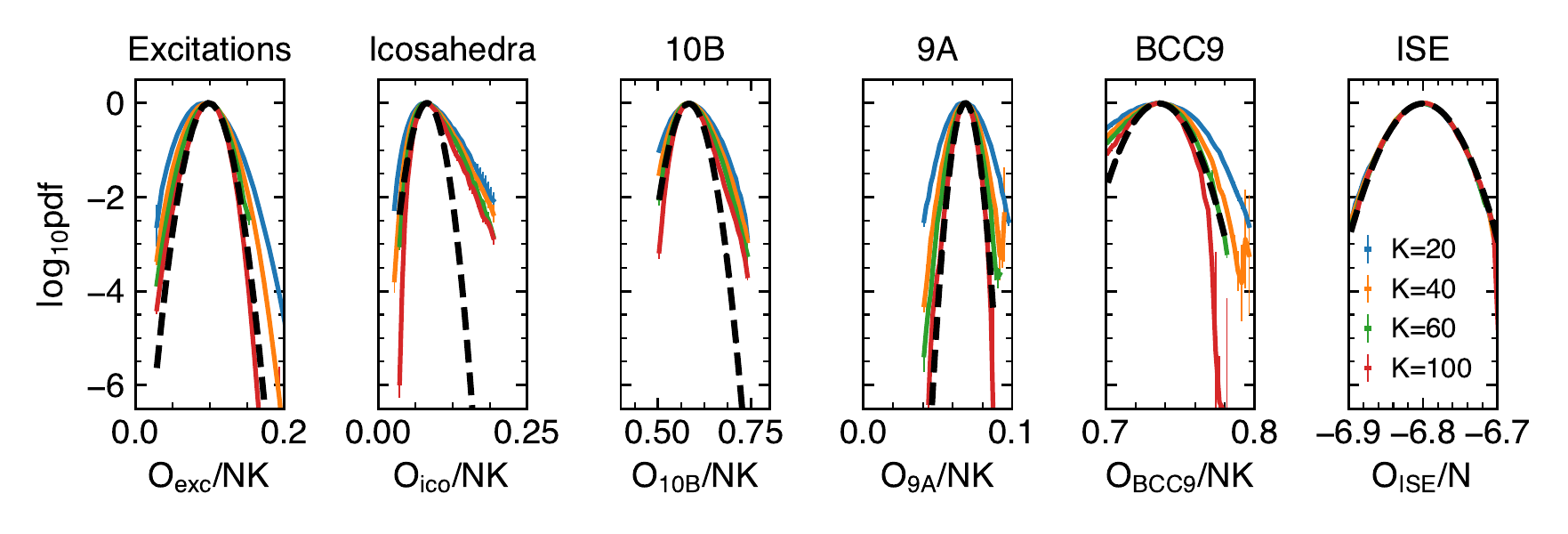}
  \caption{Probability distributions as a function of the mobility-conjugated field $s$ for dynamical trajectories of the Wahnstr\"om model of duration $K=20$ at temperature $T=0.67$. The several observables $x$ are: the time-integrated population of mobility excitations, the inherent state energy (ISE) of the central configuration of every trajectory, and the time-integrated population of particles in icosahedral motifs, 9A motifs and 10B motifs. The blue, green, yellow and red datasets refer to trajectories of length $K=20, 40,60,100$ respectively. The dashed black lines are fits to a Gaussian probability distribution centred at the peak value.}
  \label{figtable}
\end{figure*}

\begin{table*}

\centering
\resizebox{\textwidth}{!}{\begin{tabular}{l|c|c|c|c|c|c}
\hline\hline
K & \shortstack{Excitations \\ ({\footnotesize Mean, Variance, $\kex$})} &\shortstack{13A\\ ({\footnotesize Mean, Variance, $\kex$})}& \shortstack{10B\\ ({\footnotesize Mean, Variance, $\kex$})} & \shortstack{9A\\ ({\footnotesize Mean, Variance, $\kex$})} & \shortstack{BCC9\\ ({\footnotesize Mean, Variance, $\kex$})} &    \shortstack{ISE\\ ({\footnotesize Mean, Variance, $\kex$})}  \\
\hline
\hline
20 & 0.09 0.0005 -0.04  & 0.09 0.0007 0.5  & 0.6 0.002 -0.1  & 0.07 7e-05 -0.06  & 0.7 0.0003 -0.5  & -7 0.0008 0.02 \\
40 & 0.1 0.0003 -0.08  & 0.09 0.0006 0.8  & 0.6 0.001 0.1  & 0.07 4e-05 0.08  & 0.7 0.0002 -0.5  & -7 0.0008 0.003 \\
60 & 0.1 0.0003 -0.03  & 0.09 0.0005 0.9  & 0.6 0.001 0.2  & 0.07 3e-05 0.06  & 0.7 0.0002 -0.4  & -7 0.0007 -0.03 \\
100 & 0.1 0.0002 0.1  & 0.09 0.0004 2.0  & 0.6 0.0008 0.6  & 0.07 2e-05 0.07  & 0.7 0.0002 -0.3  & -7 0.0007 -0.005 \\
\hline\hline
\end{tabular}}
\caption{Expectation values for the mean, the variance and the excess kurtosis $\kex$ for several observables as measured via trajectory sampling for different values of the trajectory length at temperature $\Ts=0.67$.}
\label{tabl:scores}
\end{table*}
\subsection{Observables}
\label{sec:observables}
We analyse the emergence of phase transitions in trajectory space by monitoring a variety of observables. We perform importance sampling in trajectory space according to time-integrated observables that are either dynamical (such as the mobility excitations) or structural (a selection of geometrically different structural motifs, see Fig.~\ref{figtauIcovsT}(b)). Furthermore, in order to relate the trajectory-space picture back to the thermodynamic picture, we also monitor the inherent state energy of the selected configurations, whose statistics in the trajectory ensemble has been proven to closely reproduce the equilibrium properties. Structures are detected employing the Topological Cluster Classification algorithm and we refer to Reference \cite{malins2013tcc} for a more detailed discussion of the geometries considered here. 

In particular, for the time-integrated quantities we have:
\begin{itemize}
	\item \textit{number of excitations}: To quantify the number of mobile particles, we compute the observable
\begin{equation}
O_{\rm excitation}=\sum_{t=-L/2}^{L/2} \sum_{i=1}^N |\Theta(\Delta \bm{r}_i(t)-a )|,
\end{equation}
where $\Theta(\Delta \bm{r}_i(t)= \bm{r}_i(t)-\bm{r}_i(t-\Delta t)$ is the single particle displacement, $\Theta$ is the Heaviside function and $a$ is a scale for cage motion, here set to $a=0.3\sigma_{\rm large}$.
	\item \textit{number of particles in icosahedral motifs:} Given the important role of icosahedral order in the Wahnstr\"om mixture, we track this specific local motif along the trajectories. Additionally, we perform importance sampling according to the number of icosahedra. The corresponding time-integrated extensive structural-dynamical observable is then
 \begin{equation}
O_{\rm ico}=\sum_{t=-L/2}^{L/2} \sum_{i=1}^{N}h_{i}^{\rm ico}(t),
\label{eq:nico}
\end{equation}
where $h_{i}^{\rm ico}$ is an indicator function, which takes value 1 if a particle is found in an icosahedral environment or 0 if it is not. \ft{With a certain abuse of language, we will interchangeably refer to the \textit{population of icosahedra} or the \textit{population of particles in icosahedral motifs} when considering the intensive quantity $O_{\rm ico}/NK$.}

	\item \textit{number of particles in 9A motifs:} We compute $O_{9A}$ performing the summation as in Eq.\ref{eq:nico}, but with a different indicator function $h_{i}^{\rm 9A}$. In this case, we consider the 9A structure of the Topological Cluster Classification, which is composed of six particles combined to form three four-folded rings, surrounded by three further spindle particles on each quadrangular facet (forming a \textit{tricapped trigonal prism}). According to previous studies \cite{malins2013jcp}, we do not expect this motif to be a good predictor of structural-dynamical heterogeneity for the Wahnstr\"om mixture. However, in the case of other simple liquids dominated by five-fold symmetric local order, such as moderately polydisperse hard-sphere, 9A motifs have been shown to be complementary to local icosahedral order, becoming less frequent when the packing fraction (and the population of icosahedra) increase \cite{royall2017}.
	\item \textit{number of particles in BCC motifs:} As a further test, we compute the time-integrated observable $O_{BCC}$ considering a nine particle structure that (weakly) correlates with body centered cubic local order and anti-correlates strongly with icosahedral and five-fold symmetric order.
	\item \textit{number of particles in five-fold symmetric motifs:} Finally, to track five-fold symmetric local order that is not fully icosahedral, we consider the defective icosahedron structure 10B, composed of three interlaced pentagonal rings. This structure is characteristic of hard-sphere mixtures, and has been shown both in simulations and experiments to drive a clear structural-dynamical phase transition \cite{pinchaipat2017}.
\end{itemize}

We also measure a static observable, i.e. not time-integrated.  This is the inherent state energy (ISE) of configurations located at the centre of each trajectory, chosen in order to avoid finite-time effects on the statistics \cite{coslovich2016}. Inherent state energies are obtained minimising the potential energy of the system  for a maximum of 1000 iterations of the \textsc{FIRE} algorithm \cite{bitzek2006}.

\subsection{$s$-ensemble}

First, we consider the response of the system to a dynamical bias. This means that we collect trajectories according the observable $O_{\rm exc}$, i.e. the time-integrated number of mobility excitations. We employ the large deviation formalism and notation, and we define $s$ as the dynamical conjugate field related to the excitations, so that positive/negative values of $s$ correspond to atypically small/large densities of mobility excitations, hence the name of $s$-ensemble \cite{garrahan2007}. As we sample the mobility large deviations, we track all the other dynamical and static order parameters. 

In Fig.~\ref{figtable} and in Table \ref{tabl:scores} we summarise our findings for a particular thermostat temperature $\Ts=0.67 T_{\rm onset}\approx 1.2 T_{\rm MC}$, where $T_{\rm MC}$ and $T_{\rm onset}$ indicate the transition temperature predicted by the power-law fit to the relaxation of mode-coupling theory and the onset of the two-step relaxation dynamics respectively. In Fig.~\ref{figtable} we compare the (scaled) logarithm of the probability distributions (\textit{i.e.} the \textit{rate function)} of the considered observables $O_x$ for increasing values of the trajectory length $\tobs= 1.9,3.8, 5.7, 9.5\tau_{\alpha}$ ($K=20,40,60,100$). 
At the considered temperature, we expect to observe deviations from Gaussian fluctuations in the tails (i.e. large deviations) of the probability distributions. 
With this comparative analysis, we want to stress that the choice of the observable is non-trivial and different observables present characteristic features. 

First we notice that the population of excitations (which is the reaction coordinate along which we perform importance sampling) shows mostly Gaussian fluctuations around the mean value $\langle O_x\rangle/KN=0.096(2)$ for all the sampled trajectory lengths. 
However, the variance computed at different trajectory lengths appears to slowly converge to smaller values, with the tails of the probability distributions gradually narrowing. This indicates that very short trajectories of length $\tobs= 1.9, K=20$ are affected by finite size effects that enhance the observation of large fluctuations. 

Higher order moments converge even more slowly but point to the emergence of non-Gaussian features. For example, the excess kurtosis is negative for short trajectories and becomes mildly positive for the longest trajectories $K=100$. This underlines that even longer trajectories are needed to obtain more marked signatures of a dynamical phase transition in terms of population of excitations at the relatively high temperature considered here, with an non-negligible increase of the computational cost. Notice that it is only in the long time limit that a large deviation principle holds and rate functions converge  \cite{touchette2017}, and therefore it is only in this limit that a formal phase transition in trajectory space is expected.

\ft{Given the weak response in the mobility excitations, what signatures do we observe in the other observables measured on the same trajectories produced in the $s$-ensemble? In the following, we analyse them one by one.} 
 
For the time-integrated population of particles in icosahedral motifs, we observe that average values do not depend on $K$; however, higher order moments show a dependence on the trajectory length. The values of the excess kurtosis $\kex$ show a marked increase in non-Gaussian features of the trajectory probability distribution, as confirmed by direct inspection of the probability distribution. The excess kurtosis is positive (i.e. fat-tails) and goes approximately from 0.48 to 2.0 when the trajectory length increases from K=20 to K=100. For a comparison, notice that for a common leptokurtic distribution of positive random variables such as the Rayleigh distribution, the excess kurtosis is $\kappa_{\rm exc}=-(6\pi^2-24\pi+16)/(4-\pi)^2$ hence $\kex\approx 0.24$ showing that the distribution for the icosahedra is even more leptokurtic. Compared to the response of the mobility excitations, the time-integrated population of icosahedra provide a much stronger signature for a dynamical phase transition. In particular we observe that populations of icosahedra of order $0.2$ are only two orders of magnitude less likely than the converged typical value $\langle O_{\rm ico}\rangle/NK=0.09$, with a strong exponential tail in the probability distribution. Non-Gaussian fluctuations are therefore stronger when tracking the time-integrated population of icosahedra than in the case of excitations.

 These results are consistent with previous literature \cite{malins2013jcp,turci2017prl,hocky2014} where the role of icosahedral motifs as locally favoured structures (LFS) of the Wahnstr\"om mixture as been discussed and their strong correlation with dynamical heterogeneities measured. They also confirm the scenario originally suggested for another popular glass-former (the Kob-Andersen mixture), whereby trajectories sampled according to time-integrals of the LFS delivered stronger signatures for a dynamical transition that mobility excitations \cite{speck2012,turci2017prx}.
\begin{figure}[t]
  \centering
  \includegraphics{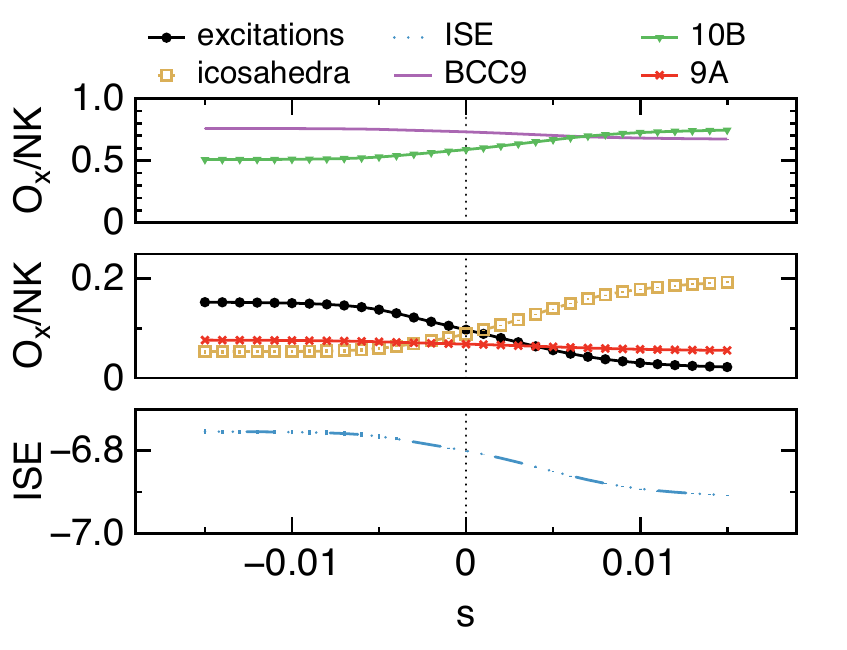}
  \caption{Dynamical transition at $\Ts=0.67$ for trajectories of length $K=60$. Excitations and icosahedra clearly anticorrelate while mobility and inherent state energies correlate. BCC9 and 9A structures are negatively correlated with the icosahedra while the five-fold symmetric 10B are positively correlated.}
  \label{figFirstMoment.pdf}
\end{figure}

An icosahedral motif is detected in the TCC via the combination of seven five-fold symmetric rings \cite{malins2013tcc}, and the statistics of the number of icosahedra appears to strongly indicate the presence of non-Gaussian fluctuations related to a structural-dynamical phase transition in the system. How does such transition change if we take into account a less restrictive observable that still identifies five-fold symmetry? To answer this question, we consider the so-called defective icosahedron structure 10B (see Sec.\ref{sec:observables} above). We first notice that the average population of particles in 10B per trajectory is much larger than the population of icosahedra (0.59 vs 0.089) and the variance again slowly converges with increasing $\tobs$. However,  The excess kurtosis is much smaller in absolute values, changing sign from negative towards positive values (leptokurtic distributions) as the trajectory length is increased.   This matches the dynamical notion of locally favoured structures: icosahedra are not only the minimum energy structure for the Wahnstr\"om interaction, they also are the individual motif (among the several options of the Topological Cluster Classification) that displays the longest persistence time \cite{malins2013jcp}. The indicators for a structural-dynamical transition in terms of 10B motifs are much weaker than in the case of icosahedral order. Yet, they confirm that the inactive (low population of excitations) regime is dominated by long-lived five-fold symmetric motifs.  

Is it possible to detect signs of the transition in other structural observables? We consider the two exemplary cases of the 9A and BCC9 structures. These motifs both correspond to arrangements of 9 particles with different symmetries which are not minimum energy clusters of the potential. The average populations of the two motifs are very different ($\sim0.07$ for 9A and $0.74$ for BCC9). The 9A probability distribution is well approximated by a Gaussian for all the trajectory lengths considered here, and the corresponding excess kurtosis are (in absolute value) the smallest among all the considered structures. The BCC9 motif, conversely, presents relatively large but negative excess kurtosis, indicating that the tails of the distributions decay more rapidly than in the case of a Gaussian distribution. 

For a given trajectory length, we consider the $s$-ensemble averages as a function of the field $s$ to highlight correlations and anticorrelations between the observables. With trajectories of length $K=60$, we show in Fig.~\ref{figFirstMoment.pdf} that for negative $s\ll0$ we sample trajectories  characterised by large densities of excitations (active phase) while for $s\gg0$ we have trajectories with low densities of excitations (inactive phase). These correspond respectively to trajectories that are poor and rich in icosahedra. The anticorrelation between mobility and five-fold symmetry is reflected also in the negative correlation between mobility and 10B structures. On the other hand, mobility positively correlates with the remaining motifs (9A and BCC9).

Finally, we consider how the active/inactive transition is translated  in terms of the energy landscape of the system. To do so we also track the inherent state energy (ISE) of the central configuration of every single trajectory and plot the corresponding probability distribution. This (as expected) does not show dependence in the trajectory length and it is well reproduced by a Gaussian fit, see Fig.~\ref{figtable}. Normal fluctuations are confirmed by the analysis of the respective excess kurtosis, which are by far the smallest measured throughout our analysis (as small as $\kex=-0.003$). In Fig.~\ref{figFirstMoment.pdf}, we do observe a transition to trajectories whose central configurations display typically much more negative energies with respect to the equilibrium typical value at $s=0$. This is consistent with the finding that in a different binary mixture (Kob-Andersen) low mobility is a good predictor of low inherent state energies \cite{coslovich2016}.

\subsection{$\mu$-ensemble}

The direct route to access structural-dynamical phase transitions is to sample trajectories according to a relevant time-integrated structural observable. From the previous discussion, and in particular from the magnitude of the non-Gaussian fluctuations as measured by the excess kurtosis, it is evident that icosahedral motifs are well suited to this purpose.

Therefore we perform additional trajectory sampling according to the time-integrated number of icosahedral motifs. As \ft{in the case of the $s$-ensemble}, we sample trajectories following the replica exchange scheme, with quadratic pseudo-potentials for the replicas with suitable spring constant $\omega$. 

In the new ensemble of trajectories, the conjugate field related to the number of particles in icosahedral motifs is termed $\mu$. Consistently with previous works in the literature \cite{speck2012,turci2017prx}, averages of any arbitrary quantity in the $\mu$-ensemble are defined as 
\begin{equation}
	\langle A\rangle_\mu = \frac{\langle A e^{\mu O_{ico}}\rangle}{\langle e^{\mu O_{ico}}\rangle}.
\end{equation}

\begin{figure}[t]
\centering
  \includegraphics{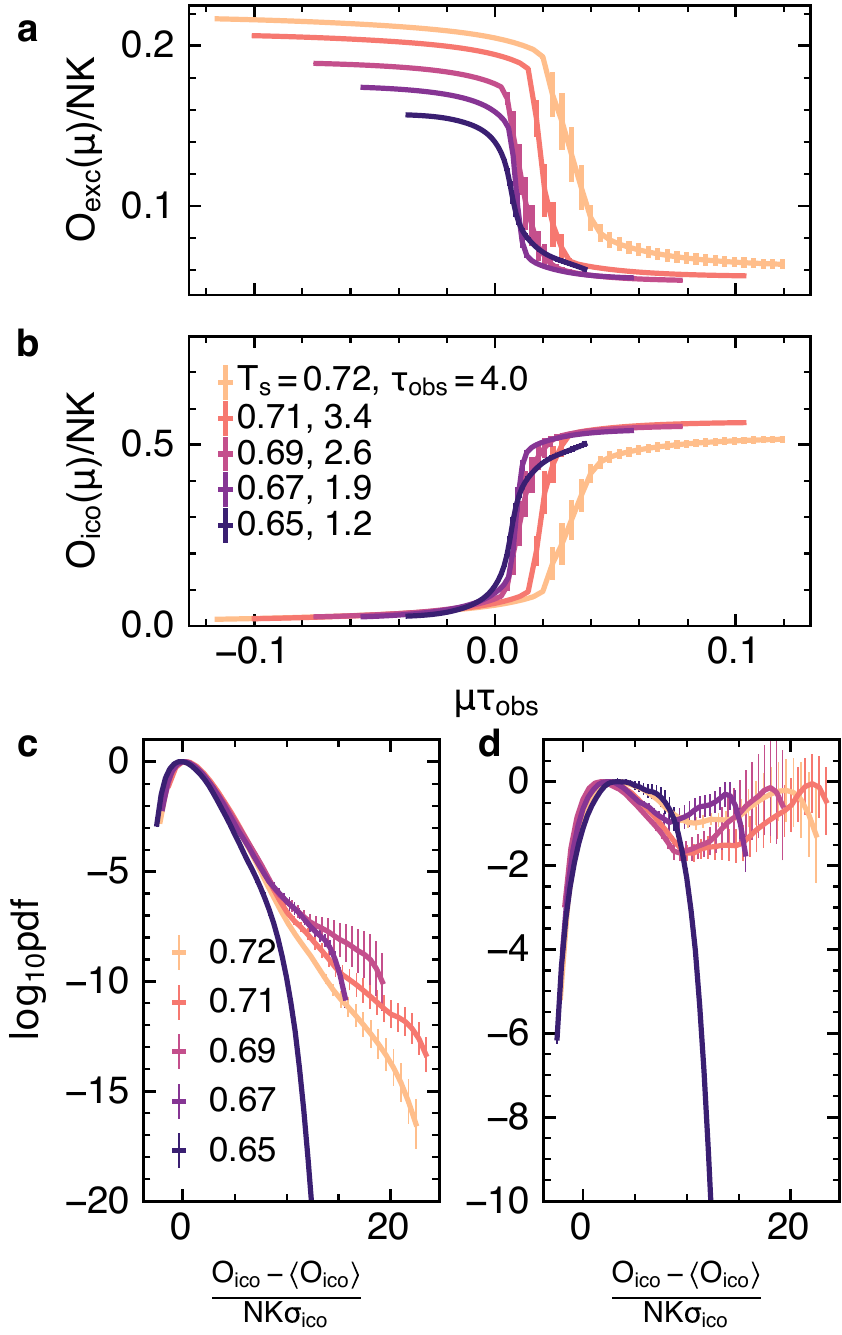}
  \caption{Structural-dynamical phase transition. (a-b) Time-integrated density of excitations (a) and of icosahedra (b) vs the rescaled conjugate field $\mtobs$ for different sampling temperatures $\Ts$. The observation time expressed in units of the structural relaxation time $\mtobs=\mu\tobs/\tau_\alpha$ is also reported. (c-d) Probability distributions of the population of icosahedra per trajectory: (c) at $\mu=0$ and (d) at $\mu=\mu^\ast(T)$, shifted by the mean value and the standard deviation $\sigma_{ico}$ in order to highlight the tail behaviour.}
  \label{fig-muensemble}
\end{figure}

In the previous section, we have shown that in the $s$-ensemble an emergent active/inactive transition is mirrored by a rapid increase of the population of particles in icosahedral motifs. In the $\mu$-ensemble we sample such structural transition directly. In Fig.~\ref{fig-muensemble}(a,b) we plot the $\mu$-dependence of the average mobility $\langle O_{\rm exc}\rangle_{\mu}/NK$ and the average population of icosahedra $\langle O_{\rm ico}\rangle_{\mu}/NK$ for several thermostat temperatures $\Ts$, from $\Ts=0.72$ to $\Ts= 0.65$. At different temperatures, we perform simulations of different trajectory lengths $\tobs$. Since the relevant time-scale for the dynamics is the structural relaxation time $\tau_{\alpha}$, we plot the first moments as a function of the non-dimensional scaled conjugate field $\mtobs:=\mu\tobs/\tau_\alpha$. Just below the onset temperature we observe signs of a phase transition at large $\mtobs$ between \ft{trajectories poor in icosahedra with high mobility and trajectories rich in icosahedra with low mobility}. As we reduce the thermostat temperature, the transition moves to values closer to $\mu=0$. Through a spline fit and the estimate of the maximum in the derivative, we obtain the value $\mstar$ at which the transition takes place. 

\begin{figure}[t]
\centering
  \includegraphics{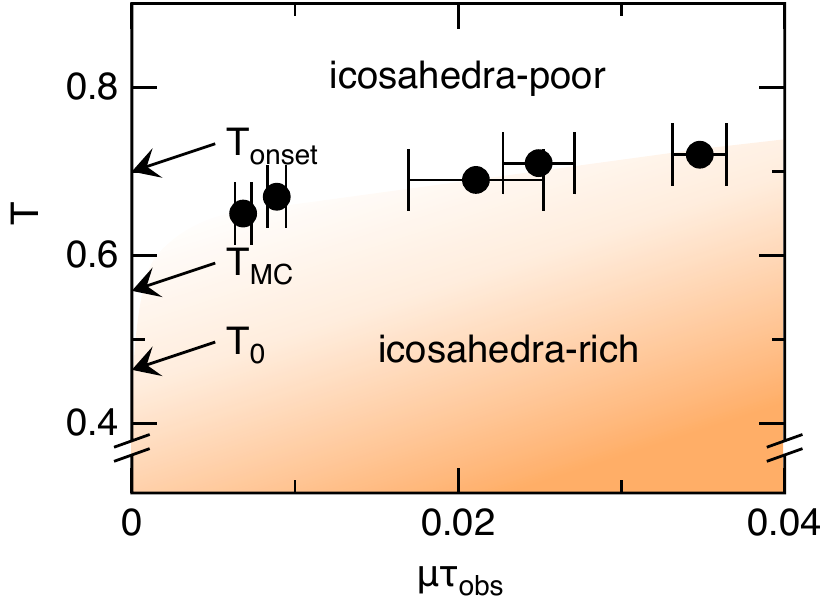}
  \caption{Structural-dynamical phase transition in the $T-\mu$ plane. The dots correspond to the couples $(\Ts, \mu^{\ast})$ determined from trajectory sampling. The blue and red dashed lines are a linear and power law fit respectively. }
  \label{figmu_T}
\end{figure}

The very small values of $\mstar$ at relatively high temperatures compared to $T_0$ obtained from the Vogel-Fulcher-Tammann fit or the mode-coupling $T_{MC}$ temperatures suggest that trajectories with an exceptionally high population of icosahedra should be highly likely, and signatures of bi-modality in the probability distribution of the time-integrated observables should become accessible even to conventional simulations as the temperature is reduced.

In Fig.~\ref{fig-muensemble}~(c,d) we plot such probability distributions both for $\mu=0$ and the critical value $\mu=\mtobs^{\ast}$, shifting and rescaling the abscissa axis by the mean and the standard deviation. We observe that, as temperature decreases, the structure-rich tail of the probability distributions raises of several orders of magnitude. Signs of bimodality are weak at low temperatures, due to the relatively short observation time $\tobs$, but clearer at higher temperatures. Moreover, if we evaluate the probability distributions at coexistence $\mu=\mu^{\ast}$, Fig.\ref{fig-muensemble}(d), a peak at high population of structures emerges more clearly. 

The knowledge of $\mstar(\Ts)$ allows us to draw an approximate structural-dynamical phase diagram, Fig.~\ref{figmu_T}, identifying the locus of points where the transition from icosahedra-poor to icosahedra-rich trajectories occurs. For the considered temperatures, we observe that most of the data points lie on a straight line. An extrapolation of the line to $\mtobs=0$ would imply that at temperature $T=0.64$ coexistence between the two structural-dynamical phases would be observable at $\mu=0$, i.e. in conventional simulations with no need for importance sampling. Previous numerical studies of the model \cite{kim2013,malins2013jcp} managed to equilibrate the supercooled liquid down to temperature $T=0.58$, with no signature of a transition while decreasing the temperature, but with a rapid increase of the population of icosahedra. This excludes the possibility of a transition at $\mu=0$ for at least $T>0.58$. As discussed in \cite{turci2017prx}, several alternative scenarios can be obtained with different extrapolations at low temperatures, including ones where the transition asymptotically reaches $\mu=0$ only in the $T\rightarrow0$ limit \cite{jack2010}. Here we notice that as we reduce the temperature, the critical field $\mstar$ is reduced by progressively smaller amounts for the successive temperatures. Lower temperature sampling is partly hindered by the long convergence times of the Monte-Carlo in trajectory space, see Fig.~\ref{figmc}.

\begin{figure}[b]
  \centering
  \includegraphics{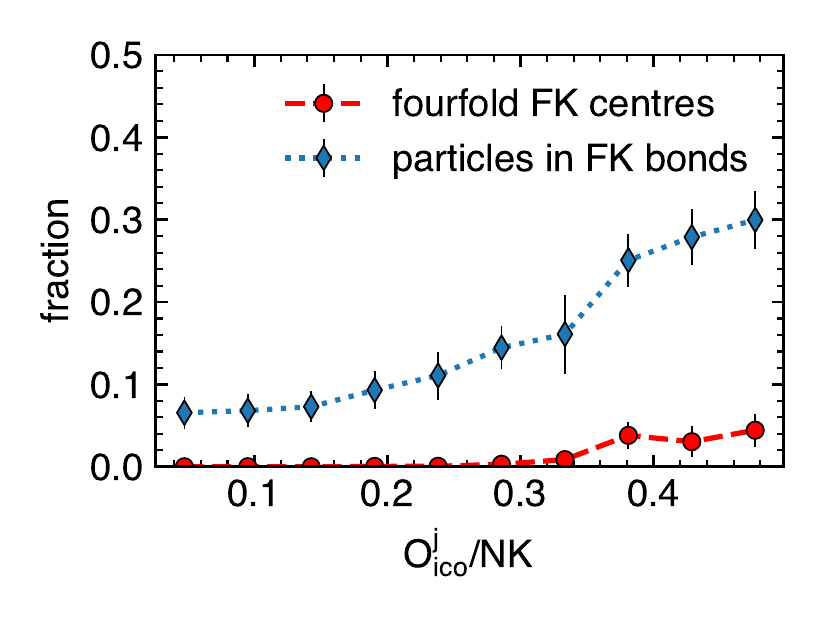}
  \caption{Analysis in terms of Frank-Kasper (FK) bonds for a representative temperature $\Ts=0.67$. The fraction of particles involved in FK bonds increases as we consider $\mu$-ensemble replicas with larger typical values of the concentration of icosahedra $O_{\rm ico}^j/NK$ (orange diamonds). However, the fraction of large particles involved in four-fold FK bonds (necessary for crystallisation) is below 5\% even for the replicas with the largest $O_{\rm ico}^j$ (blue circles). Errorbars indicate standard deviations and lines are guides to the eye. }
  \label{figFK}
\end{figure}

In the icosahedra-rich regime, approximately 50\% of the particles can be found in a local icosahedral environment. However, a complex unit cell formed by several icosahedra and Frank-Kasper bonds has been shown to drive the system towards crystallisation \cite{pedersen2010}. We check this possibility by monitoring the concentration of Frank-Kasper bonds, here defined as pairs of large A particles surrounded by six common B particles. In Fig.~\ref{figFK} we plot the average fraction of particles involved in Frank-Kasper bonds for the increasing reference concentration of icosahedra $O_{\rm ico}^j$ in the replica-exchange scheme at an exemplary temperature $\Ts=0.67$. We observe a rapid increase in the number of Frank-Kasper bonds as we consider replicas with very high concentrations of icosahedra. This is consistent with the overall behaviour of the Wahnstr\"om supercooled liquid at low temperatures, where Frank-Kasper bonds are very common \cite{pedersen2010}. However, in order to form a crystalline phase, four-fold Frank-Kasper bonds between the large particle species are necessary. If we focus on the fraction of A particles in four-fold bonds, this increases very mildly across all of the replicas, and stays below 5\% in the highest bias replica, excluding crystal formation in the icosahedra-rich phase.
\newline

In conclusion, both the $s$ and the $\mu$ ensemble calculations provide evidence for an inactive \textit{and} icosahedra-rich dynamical phase that becomes progressively more likely to be observed  for $T<T_{\rm onset}$. We now study the icosahedral phase more in detail to understand its relation with the emergence of rigidity in the glass.


\section{Rheological response of the inactive/icosahedra-rich phase}
\label{sec:shear}
As a supercooled liquid is cooled down, it eventually undergoes an experimental glass transition where the relaxation time $\tau_\alpha$ exceeds the available observation time by many orders of magnitude. Such a phenomenological transition is accompanied by the emergence of solid behaviour: the glass behaves like a solid, in the sense that it can be probed through rheological measurements, proving a finite elastic response and shear modulus.

We have shown that as the temperature is decreased, the Wahnstr\"om mixture explores more and more frequently trajectories that are exceptionally rich in structure. Moreover, the icosahedra-rich trajectories not only are characterised by low mobility (inactive trajectories) but they also tend to have configurations with low inherent state energies. Is it possible to connect these structural and dynamical changes to the emergence of  solidity, i.e. to the rheological response of the system?

We test this idea realising an ensemble of configurations extracted from the trajectories produced in the $\mu$ ensemble at the thermostat temperature $\Ts=0.65$. From every umbrella $i$ of the replica-exchange algorithm we extract a population of configurations that are representative of the fluctuations, in trajectory space, around a specific value of the population of icosahedra 
\begin{equation}
n_i^0=O_{\rm ico}/NK.
\end{equation}

In particular, we produce a discrete group of 8 sets with 75 initial configurations each at the following typical population of icosahedra $n_i^0=[0.09, 0.17,0.29,0.33,0.38,0.43,0.49,0.55]$. According to the available data and the extrapolation of the fit shown in Fig.~\ref{figtauIcovsT}, these populations would be typical in the equilibrium supercooled at temperatures [0.67, 0.60, 0.54, 0.52, 0.50, 0.49, 0.46]. In our simulations, we take averages for every set of initial conditions extracted from distinct replicas.

To understand the purely mechanical response of the different sets of configurations we study the linear shearing of the system in the Athermal Quasi-static limit (AQS) \cite{maloney2004,maloney2006}. Under this protocol, the system is slowly deformed in a chosen direction at a fixed shear rate $\dot{\gamma}=0.005$ for a small time interval $\Delta t=0.005$; subsequently, the \textsc{FIRE} energy minimisation algorithm \cite{bitzek2006} is employed to lead the particles to the closest inherent state. The two steps are repeated until the system reaches a maximum total strain of $\gamma=0.5$

\begin{figure}[t]
  \centering
  \includegraphics{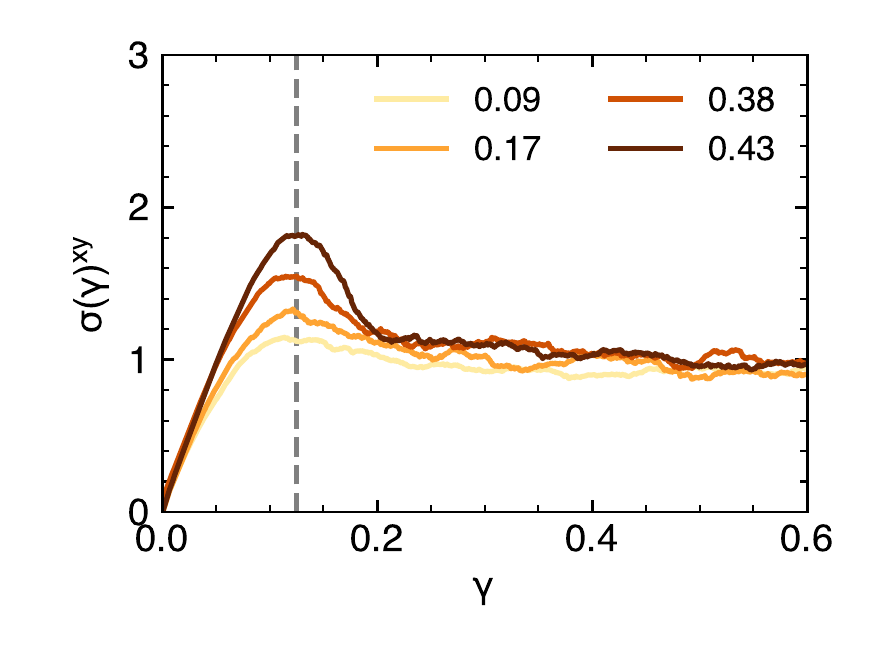}
  \caption{Average shear-stress as a function of shear strain for different values of the typical initial population of icosahedra in $\mu$-ensemble configurations $n_i^0$. The vertical dashed line corresponds to the yield strain. }
  \label{figAQSstress}
\end{figure}

\begin{figure}[t]
  \centering
  \includegraphics{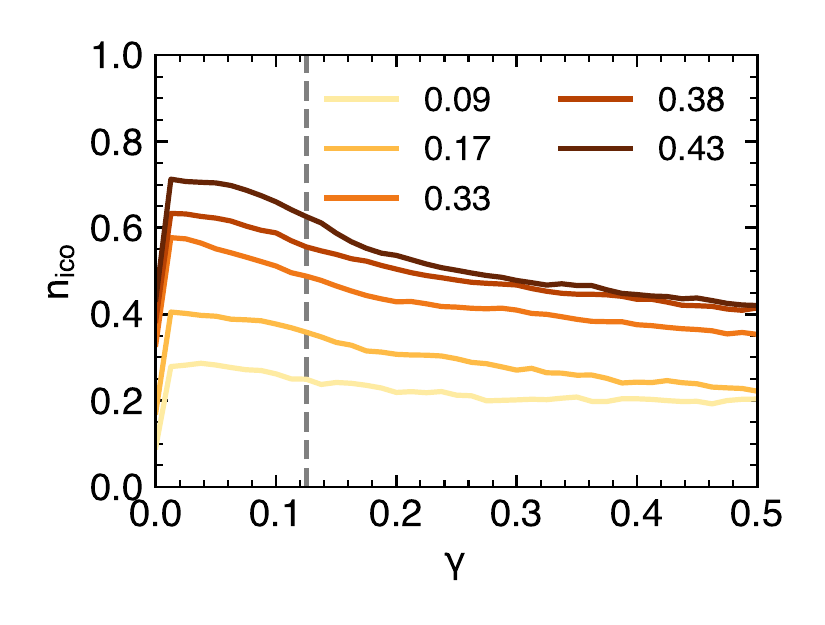}
  \caption{Overall average population of icosahedra as a function of shear strain for different values of the typical initial population of icosahedra in $\mu$-ensemble configurations $n_i^0$. The vertical dashed line corresponds to the yield strain.}
  \label{figNicoShear}
\end{figure}

In Fig.~\ref{figAQSstress} and Fig.~\ref{figNicoShear} we plot the response of the system in terms of shear stress $\sigma^{xy}$ and fraction of particles in icosahedral domains for different typical values of the initial population of icosahedra $n_i^0$. The first striking result is that the yield stress $\sigma_{\rm yield} = \max_\gamma \sigma^{xy}(\gamma)$ strongly depends on  $n_i^0$, and it approximately doubles as the typical population of icosahedra quadruples. The yield strain $\gamma_{\rm yield}$ (the value of strain at which the maximum stress is reached) is not sensitive to the different starting conditions and is located at approximately $\gamma_{\rm yield}\approx0.12$ for the chosen strain rate. At the same time, we notice that the shear protocol induces a sudden increase of the population of icosahedra at very early times (very small strains) and a progressive decay of the population which accelerates as the yield strain is reached. The overall, instantaneous increase of the population of icosahedra can be understood as a consequence of the minimisation procedure, which destroys thermal fluctuations present in the initial configurations and promotes the formation of local minimum energy motifs, such as the icosahedron. This implies that the overall population of icosahedra $n_{\rm ico}(\gamma)$ can be distinguished into two families: the first refers to the subset of particles that are located in icosahedral domains in the original starting configurations produced in the $\mu$ ensemble, and it is identified by the boolean vector $\etax{\mu}$ of length $N$; the second refers to all the remaining particles in icosahedral domains, resulting from the AQS protocol, identified by the vector $\etax{\bar{\mu}}$.

As the system is sheared, the number of icosahedra changes very mildly for strains below the yield strain, and only later declines, supporting the idea that the population of icosahedra is related to the rigid, elastic response of the system. Having defined two subpopulations of icosahedra, we now quantify their respective differences in the mechanical response.

\begin{figure}[t]
  \centering
  \includegraphics{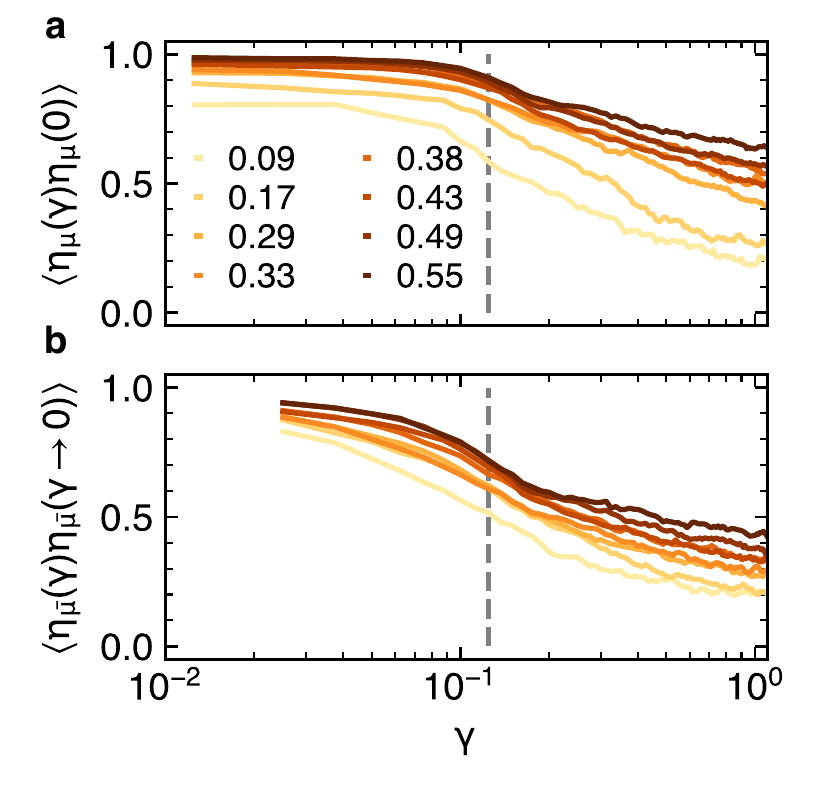}
  \caption{Auto-correlation of the probability for a particle to be in icosahedral domain for the $\eta_\mu$ and the $\eta_{\bar{\mu}}$ populations (see main text for definition) for different values of the typical initial population of icosahedra.The $\etax{\bar{\mu}}$ family is not defined at $\gamma=0$ so, we take the smallest $\gamma$ as the reference state. The vertical dashed line corresponds to the yield strain.
  }
  \label{figCorssshear}
\end{figure}

To do so, we compute separate auto-correlation functions $\langle\etax{x}(\gamma)\etax{x}(0)\rangle$ for the $\etax{\mu}$ and the $\etax{\bar{\mu}}$ populations, Fig.~\ref{figCorssshear}(a,b). We notice that only for large initial populations of icosahedra the autocorrelation functions start close to unity. This shows that the reorganisation induced by the AQS protocol not only forms new icosahedral motifs, but it also initially destroys a fraction of them. The two families of autocorrelation functions show distinctively different behaviours: the icosahedra present in the initial $\mu$-ensemble, Fig.~\ref{figCorssshear}(a), show a long plateau that terminates only when the yield strain is attained; the icosahedra generated via AQS, Fig.~\ref{figCorssshear}(b), continuously decorrelate at earlier times (smaller strains).

\begin{figure}[t]
  \centering
  \includegraphics{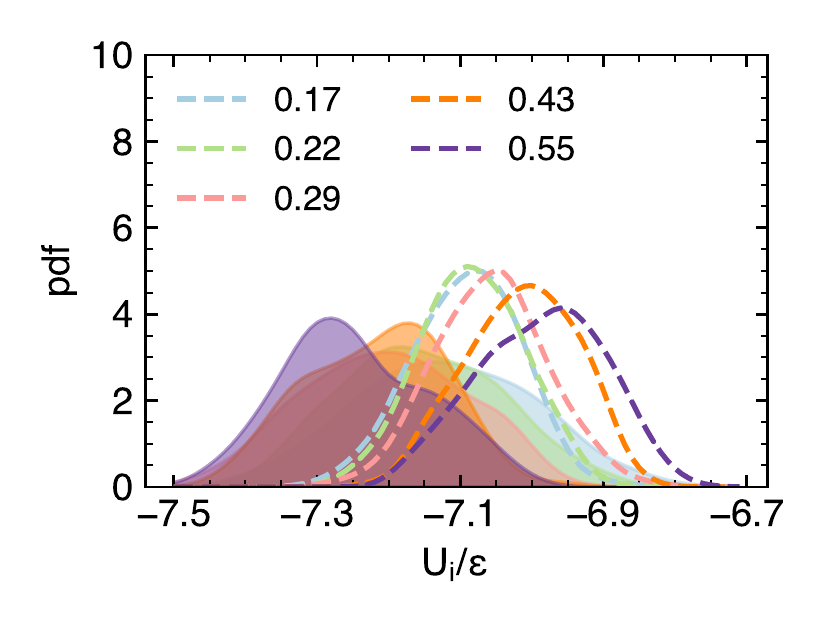}
  \caption{Distribution of potential energies for particles belonging to the $\eta_\mu$ (filled distributions) and the $\eta_{\bar{\mu}}$ (dashed curves) families (see main text for definition). The distributions are computed over the whole duration of the shearing protocol and averaged over 75 initial configurations. }
  \label{figSmoothedhists}
\end{figure}

A further confirmation of the different responses between the $\etax{\mu}$ and the $\etax{\bar{\mu}}$ icosahedra is provided by the distribution of the potential energy of the individual particles constituting the two families. In Fig.~\ref{figSmoothedhists} we plot the overall energy distributions, for the population, collected all along the shearing protocol. We clearly observe that not all icosahedral motifs are energetically the same: particles located in icosahedral motifs purely emerging from energy minimisation  have energies that are typically higher than particles identified in icosahedral motifs in the original $\mu$ ensemble configurations. The energy gap between the two families widens as we consider initial configurations with larger concentrations of icosahedra: at very high initial concentrations (55\%), the $\etax{\mu}$ subpopulation has energies that are 8\% lower than the $\etax{\bar{\mu}}$ subpopulation.

%
%
%
%
%
%
%
%
%
%
%

\section{Conclusions}

Through numerical simulations, we have discussed a third example of structural-dynamical phase transition in a model of atomistic glassformer, after the previously considered cases of the Kob-Andersen mixture \cite{hedges2009,jack2011,speck2012jcp,speck2012,turci2017prx} and the moderately polydisperse hard-spheres \cite{pinchaipat2017}.

A quantitative analysis of the probability distributions of time-integrated observables demonstrates that well-chosen time-integrated structural motifs can be used to perform efficient importance sampling. In particular, it makes possible to explore structure-rich trajectories (representative of colder temperature states \cite{turci2017prx}) that are otherwise hard to reach. At the same time, we find confirmation for a sharp (first-order) transition in trajectory space, that becomes measurable below the onset temperature, between a structure-rich and a structure poor dynamical phase, the former becoming more and more likely as the temperature is reduced, similarly to what has been previously observed in the Kob-Andersen mixture \cite{turci2017prx}. Within the range of the explored temperatures, it is impossible to assess what the low temperature fate of the transition may be: the reduction of the critical conjugate field value $\mu^{\ast}$ suggests that, as the temperature decreases, the structure-rich phase would prevail. However, it is unclear whether the previously reported  crystallisation into complex Laves phases \cite{pedersen2010} would interfere with the emergence of the icosahedra-rich phase. In our simulations, we monitored the evolution Frank-Kasper bonds (an essential element of the complex crystalline phase) and do not find a significative increase in the icosahedra-rich phase compared to the icosahedra-poor phase.

In Reference \cite{turci2017prx}, the study of the alternative Kob-Andersen mixture at low temperature indicated possible scenarios for the temperature dependence of the structural-dynamical transition. In the present case of the Wahnstr\"om mixture, we observe that at $0.72 T_{\rm onset}$ the structure-rich phase is highly metastable, while a relatively modest decrease of the temperature to $0.67 T_{\rm onset}$ makes the exploration of the structure-rich basin 5 to 7 times more likely, see Fig.~\ref{fig-muensemble}~(d). The temperature dependence of the critical value $\mu^\ast\tau_{\rm obs}$ shows a decrease towards $\mu=0$ which becomes less pronounced as the temperature is decreased. This is accompanied by strong correlations between successive steps in the trajectory-space Monte-Carlo that slow down equilibration and make lower temperature sampling particularly challenging. The present data support the narrowing of the free-energy gap (in trajectory space) between the structure-rich and structure-poor states when the temperature is decreased, and do not exclude the possibility that the transition terminates at a lower critical point at finite temperature, as in kinetically constrained models with additional softness \cite{elmatad2010}.

In order to better understand the importance of the structure-rich phase, configurations obtained through trajectory sampling have been probed through an out-of-equilibrium rheological protocol, effectively regarding these configurations as samples of an amorphous material at $T=0$. Consistently with previous studies of the Wahnstr\"om mixture based on conventional simulations \cite{pinney2016}, we find that icosahedra play a major role in the emergence of rigidity: icosahedra-rich configurations display much larger yield stresses than icosahedra-poor ones. However, we nuance this statement, as we are able to split the overall family of icosahedral motifs according to the preparation protocol: well-thermalised configurations from trajectory sampling have icosahedral regions that are more robust to shear and with lower energies than the icosahedral domains obtained via energy minimisation. This highlights that the requirement of sampling long-lived structural motifs (implicit in trajectory sampling) allows us to explore metabasins that are not just richer in structure, but more stable as well.

\acknowledgement
 CPR acknowledges the Royal Society for funding. FT and CPR acknowledge the European Research Council (ERC consolidator grant NANOPRS, project number 617266). This work was carried out using the computational facilities of the Advanced Computing Research Centre, University of Bristol. FT contributed to the generation and analysis of the data and the writing of the manuscript. FT, TS and CPR contributed to the editing of the manuscript.

%
%
\bibliographystyle{unsrt}
\bibliography{allDropBox}

\begin{thebibliography}{10}

\bibitem{cavagna2009}
A.~Cavagna.
\newblock Supercooled liquids for pedestrians.
\newblock {\em Phys. Rep.}, 476:51--124, 2009.

\bibitem{berthier2011}
L.~Berthier and G.~Biroli.
\newblock Theoretical perspective on the glass transition and amorphous
  materials.
\newblock {\em Rev. Mod. Phys.}, 83:587--645, 2011.

\bibitem{royall2015physrep}
C.~P. Royall and S.~R. Williams.
\newblock The role of local structure in dynamical arrest.
\newblock {\em Phys. Rep.}, 560:1, 2015.

\bibitem{barrat2011}
J-L. Barrat and A.~Lema\^{i}tre.
\newblock {\em Dynamical heterogeneities in glasses, colloids, and granular
  media}, chapter Heterogeneities in amorphous systems under shear.
\newblock Oxford University Press, 2011.

\bibitem{torquato2000}
S.~Torquato, T.~M. Truskett, and P.~G. Debenedetti.
\newblock Is random close packing of spheres well defined?
\newblock {\em Phys. Rev. Lett.}, 84(10):2064--2067, March 2000.

\bibitem{montanari2006}
A.~Montanari and G.~Semerjian.
\newblock Rigorous inequalities between length and time scales in glassy
  systems.
\newblock {\em J. Stat. Phys.}, 125:23--54, 2006.

\bibitem{hocky2014pre}
G.~M. Hocky, L.~Berthier, W.~Kob, and D.~Reichman.
\newblock Crossovers in the dynamics of supercooled liquids probed by an
  amorphous wall.
\newblock {\em Phys. Rev. E}, 89:052311, 2014.

\bibitem{royall2015}
C.~P. Royall, A.~Malins, A.~J. Dunleavy, and R.~Pinney.
\newblock Strong geometric frustration in model glassformers.
\newblock {\em J. Non-Cryst. Solids}, 407:34--43, 2015.

\bibitem{royall2008}
C.~P. Royall, S.~R. Williams, T.~Ohtsuka, and H.~Tanaka.
\newblock Direct observation of a local structural mechanism for dynamic
  arrest.
\newblock {\em Nature Mater.}, 7:556, 2008.

\bibitem{wochner2009}
Peter Wochner, Christian Gutt, Tina Autenrieth, Thomas Demmer, Volodymyr
  Bugaev, Alejandro~D{\'\i}az Ortiz, Agn{\`e}s Duri, Federico Zontone, Gerhard
  Gr{\"u}bel, and Helmut Dosch.
\newblock X-ray cross correlation analysis uncovers hidden local symmetries in
  disordered matter.
\newblock {\em Proceedings of the National Academy of Sciences},
  106(28):11511--11514, 2009.

\bibitem{leocmach2012}
M.~Leocmach and H.~Tanaka.
\newblock Roles of icosahedral and crystal-like order in the hard spheres glass
  transition.
\newblock {\em Nature Comm.}, 3:974, 2012.

\bibitem{royall2018jcp}
C~Patrick Royall, Stephen~R Williams, and Hajime Tanaka.
\newblock Vitrification and gelation in sticky spheres.
\newblock {\em The Journal of chemical physics}, 148(4):044501, 2018.

\bibitem{sheng2006}
HW~Sheng, WK~Luo, FM~Alamgir, JM~Bai, and E~Ma.
\newblock Atomic packing and short-to-medium-range order in metallic glasses.
\newblock {\em Nature}, 439(7075):419, 2006.

\bibitem{chen2011}
C.~Chen, O.~Cook, C.~E. Nicholson, and S.~J. Cooper.
\newblock Leapfrogging ostwald's rule of stages: Crystallization of stable
  g-glycine directly from microemulsions 2011, 11,.
\newblock {\em Crystal Growth and Design}, 11:2228--2237., 2011.

\bibitem{hirata2013}
A.~Hirata, L.~J. Kang, T.~Fujita, B.~Klumov, K.~Matsue, M.~Kotani, A.~R.
  Yavari, and M.~W. Chen.
\newblock Geometric frustration of icosahedron in metallic glasses.
\newblock {\em Science}, 341:376---379, 2013.

\bibitem{gokhale2014}
S.~Gokhale, K.~H. Nagamanasa, R.~Ganapathy, and A.~K. Sood.
\newblock Growing dynamical facilitation on approaching the random pinning
  colloidal glass transition.
\newblock {\em Nature Comm.}, 5:4685, 2014.

\bibitem{chikkadi2014}
V~Chikkadi, D~M Miedema, M~T Dang, B~Nienhuis, and P~Schall.
\newblock Shear banding of colloidal glasses: Observation of a dynamic
  first-order transition.
\newblock {\em Phys. Rev. Lett.}, 113(20):208301, 2014.

\bibitem{bernal1959}
J.~D. Bernal.
\newblock A geometrical approach to the structure of liquids.
\newblock {\em Nature}, 183:141--147, 1959.

\bibitem{bernal1960}
J.~D. Bernal.
\newblock Geometry of the structure of monatomic liquids.
\newblock {\em Nature}, 185:68--70, 1960.

\bibitem{finney1970}
J.~L. Finney.
\newblock Random packings and the structure of simple liquids. i. the geometry
  of random close packings.
\newblock {\em Proc. R. Soc. A}, 319:479--493, 1970.

\bibitem{finney1970mc}
J.~L. Finney.
\newblock Random packings and the structure of simple liquids. ii. the
  molecular geometry of simple liquids.
\newblock {\em Proc. R. Soc. A}, 319:495--507, 1970.

\bibitem{nelson1983}
D.~R. Nelson.
\newblock Liquids and glasses in spaces of incommensurate curvature.
\newblock {\em Phys. Rev. Lett.}, 50(13):982--985, mar 1983.

\bibitem{tarjus2005}
G.~Tarjus, S~.A. Kivelson, Z.~Nussinov, and P.~Viot.
\newblock The frustration-based approach of supercooled liquids and the glass
  transition: a review and critical assessment.
\newblock {\em J. Phys.: Condens. Matter}, 17:R1143--R1182, 2005.

\bibitem{chandler2010}
David Chandler and Juan~P Garrahan.
\newblock {Dynamics on the way to forming glass: bubbles in space-time.}
\newblock {\em Annu. Rev. Condens. Matt. Phys.}, 61:191--217, January 2010.

\bibitem{dyre2006}
J.~C. Dyre.
\newblock Colloquium: The glass transition and elastic models of glass-forming
  liquids.
\newblock {\em Rev. Mod. Phys.}, 78:953--972, 2006.

\bibitem{lubchenko2007}
V.~Lubchenko and P.~Wolynes.
\newblock Theory of structural glasses and supercooled liquids.
\newblock {\em Annu. Rev. Phys. Chem.}, 58:235--266, 2007.

\bibitem{lubchenko2015theory}
Vassiliy Lubchenko.
\newblock Theory of the structural glass transition: A pedagogical review.
\newblock {\em Advances in Physics}, 64(3):283--443, 2015.

\bibitem{berthier2016}
L.~Berthier and M.~D. Ediger.
\newblock Facets of glass physics.
\newblock {\em Phys. Today}, 69:40--46, 2016.

\bibitem{cammarota2012}
C.~Cammarota and G.~Biroli.
\newblock {Ideal glass transitions by random pinning.}
\newblock {\em Proc. Nat. Acad. Sci}, 109(23):8850--5, June 2012.

\bibitem{kob2013}
W.~Kob and L.~Berthier.
\newblock Probing a liquid to glass transition in equilibrium.
\newblock {\em Phys. Rev. Lett.}, 110:245702, 2013.

\bibitem{kob2014nonlinear}
Walter Kob and Daniele Coslovich.
\newblock Nonlinear dynamic response of glass-forming liquids to random
  pinning.
\newblock {\em Physical Review E}, 90(5):052305, 2014.

\bibitem{berthier2016prl}
Ludovic Berthier, Daniele Coslovich, Andrea Ninarello, and Misaki Ozawa.
\newblock Equilibrium sampling of hard spheres up to the jamming density and
  beyond.
\newblock {\em Phys. Rev. Lett.}, 116(23):238002, 2016.

\bibitem{ninarello2017}
A.~Ninarello, L.~Berthier, and D.~Coslovich.
\newblock Models and algorithms for the next generation of glass transition
  studies.
\newblock {\em Phys. Rev. X}, 7:021039, 2017.

\bibitem{hedges2009}
L.~O. Hedges, R.~L. Jack, J.~P. Garrahan, and D.~Chandler.
\newblock Dynamic order-disorder in atomistic models of structural glass
  formers.
\newblock {\em Science}, 323:1309--1313, 2009.

\bibitem{speck2012}
T.~Speck, A.~Malins, and C.~P. Royall.
\newblock First-order phase transition in a model glass former: Coupling of
  local structure and dynamics.
\newblock {\em Phys. Rev. Lett.}, 109:195703, 2012.

\bibitem{turci2017prx}
F~Turci, C~P Royall, and T~Speck.
\newblock Nonequilibrium phase transition in an atomistic glassformer: The
  connection to thermodynamics.
\newblock {\em Physical Review X}, 7(3):031028, 2017.

\bibitem{garrahan2007}
J.~P. Garrahan, R.~L. Jack, E.~Lecomte, V. amd~Pitard, K.~van Duijvendijk, and
  F.~van Wijland.
\newblock Dynamical first-order phase transition in kinetically constrained
  models of glasses.
\newblock {\em Phys. Rev. Lett.}, 98:195702, 2007.

\bibitem{keys2011}
A.~S. Keys, L.~O. Hedges, J.~P. Garrahan, S.~C. Glotzer, and D.~Chandler.
\newblock Excitations are localized and relaxation is hierarchical in
  glass-forming liquids.
\newblock {\em Phys. Rev. X}, 1:021013, 2011.

\bibitem{elmatad2009}
Y.~S. Elmatad, D.~Chandler, and J.~P. Garrahan.
\newblock Corresponding states of structural glass formers.
\newblock {\em J. Phys. Chem. B}, 113:5563--5567, 2009.

\bibitem{turci2011}
Francesco Turci and Estelle Pitard.
\newblock Large deviations and heterogeneities in a driven kinetically
  constrained model.
\newblock {\em EPL (Europhysics Letters)}, 94(1):10003, 2011.

\bibitem{nemoto2017}
Takahiro Nemoto, Robert~L Jack, and Vivien Lecomte.
\newblock Finite-size scaling of a first-order dynamical phase transition:
  Adaptive population dynamics and an effective model.
\newblock {\em Physical review letters}, 118(11):115702, 2017.

\bibitem{jack2011}
R.~L. Jack, L.~O. Hedges, J.~P. Garrahan, and D.~Chandler.
\newblock Preparation and relaxation of very stable glassy states of a
  simulated liquid.
\newblock {\em Phys. Rev. Lett.}, 107:275702, 2011.

\bibitem{pitard2011dynamic}
Estelle Pitard, Vivien Lecomte, and Fr{\'e}d{\'e}ric Van~Wijland.
\newblock Dynamic transition in an atomic glass former: A molecular-dynamics
  evidence.
\newblock {\em EPL (Europhysics Letters)}, 96(5):56002, 2011.

\bibitem{isobe2016}
Masaharu Isobe, Aaron~S Keys, David Chandler, and Juan~P Garrahan.
\newblock Applicability of dynamic facilitation theory to binary hard disk
  systems.
\newblock {\em Physical review letters}, 117(14):145701, 2016.

\bibitem{speck2012jcp}
T.~Speck and D.~Chandler.
\newblock Constrained dynamics of localized excitations causes a
  non-equilibrium phase transition in an atomistic model of glass formers.
\newblock {\em J. Chem. Phys.}, 136:184509, 2012.

\bibitem{pinchaipat2017}
Rattachai Pinchaipat, Matteo Campo, Francesco Turci, James~E Hallett, Thomas
  Speck, and C~Patrick Royall.
\newblock Experimental evidence for a structural-dynamical transition in
  trajectory space.
\newblock {\em Physical review letters}, 119(2):028004, 2017.

\bibitem{hocky2014}
G.~M. Hocky, D.~Coslovich, A.~Ikeda, and D.~Reichman.
\newblock Correlation of local order with particle mobility in supercooled
  liquids is highly system dependent.
\newblock {\em Phys. Rev. Lett.}, 113:157801, 2014.

\bibitem{wahnstrom1991}
G\"{o}ran Wahnstr\"{o}m.
\newblock {Molecular-dynamics study of a supercooled two-component
  Lennard-Jones system}.
\newblock {\em Phys. Rev. A}, 44:3752--3764, Sep 1991.

\bibitem{malins2013jcp}
A.~Malins, J.~Eggers, C.~P. Royall, S.~R. Williams, and H.~Tanaka.
\newblock Identification of long-lived clusters and their link to slow dynamics
  in a model glass former.
\newblock {\em J. Chem. Phys.}, 138:12A535, 2013.

\bibitem{pinney2015}
R.~Pinney, T.~B. Liverpool, and Royall~C. P.
\newblock Recasting a model atomistic glassformer as a system of icosahedra.
\newblock {\em J. Chem. Phys.}, 143:244507, 2015.

\bibitem{coslovich2011}
Daniele Coslovich.
\newblock {Locally preferred structures and many-body static correlations in
  viscous liquids}.
\newblock {\em Phys. Rev. E}, 83(5):8, May 2011.

\bibitem{coslovich2007}
D.~Coslovich and G.~Pastore.
\newblock Understanding fragility in supercooled lennard-jones mixtures. i.
  locally preferred structures.
\newblock {\em J. Chem. Phys}, 127:124504, 2007.

\bibitem{turci2017prl}
F~Turci, G~Tarjus, and C~P Royall.
\newblock From glass formation to icosahedral ordering by curving
  three-dimensional space.
\newblock {\em Physical Review Letters}, 118(21):215501, 2017.

\bibitem{pedersen2010}
U.~R. Pedersen, T.~B. Schroder, J.~C. Dyre, and P.~Harrowell.
\newblock Geometry of slow structural fluctuations in a supercooled binary
  alloy.
\newblock {\em Phys. Rev. Lett.}, 104:105701, 2010.

\bibitem{bolhuis2002}
P.~G. Bolhuis, D.~Chandler, C.~Dellago, and P.~L. Geissler.
\newblock Transition path sampling: Throwing ropes over rough mountain passes,
  in the dark.
\newblock {\em Annual review of physical chemistry}, 53:291--318, 2002.

\bibitem{jenkinson2017}
Thomas Jenkinson, Peter Crowther, Francesco Turci, and C~Patrick Royall.
\newblock Weak temperature dependence of ageing of structural properties in
  atomistic model glassformers.
\newblock {\em The Journal of chemical physics}, 147(5):054501, 2017.

\bibitem{minh2009}
David~DL Minh and John~D Chodera.
\newblock Optimal estimators and asymptotic variances for nonequilibrium
  path-ensemble averages.
\newblock {\em The Journal of chemical physics}, 131(13):134110, 2009.

\bibitem{malins2013tcc}
A.~Malins, S.~R. Williams, J.~Eggers, and C.~P. Royall.
\newblock Identification of structure in condensed matter with the topological
  cluster classification.
\newblock {\em J. Chem. Phys.}, 139:234506, 2013.

\bibitem{royall2017}
C.~P. Royall and W.~Kob.
\newblock Locally favoured structures and dynamic length scales in a simple
  glass-former.
\newblock {\em J. Stat. Mech.: Theory and Experiment}, page 024001, 2017.

\bibitem{coslovich2016}
D.~Coslovich and R.~L. Jack.
\newblock Structure of inactive states of a binary lennard-jones mixture.
\newblock {\em J. Stat. Mech.: Theory and Experiment}, page 074012, 2016.

\bibitem{bitzek2006}
Erik Bitzek, Pekka Koskinen, Franz G{\"a}hler, Michael Moseler, and Peter
  Gumbsch.
\newblock Structural relaxation made simple.
\newblock {\em Physical review letters}, 97(17):170201, 2006.

\bibitem{touchette2017}
Hugo Touchette.
\newblock Introduction to dynamical large deviations of markov processes.
\newblock {\em Physica A: Statistical Mechanics and its Applications}, 2017.

\bibitem{kim2013}
K.~Kim and S.~Saito.
\newblock Multiple length and time scales of dynamic heterogeneities in model
  glass-forming liquids: A systematic analysis of multi-point and multi-time
  correlations.
\newblock {\em J. Chem. Phys.}, 138:12A506, 2013.

\bibitem{jack2010}
R.~L. Jack and J.~P. Garrahan.
\newblock Metastable states and space-time phase transitions in a spin-glass
  model.
\newblock {\em Phys. Rev. E}, 81:011111, 2010.

\bibitem{maloney2004}
C.~Maloney and A.~Lema\^{i}tre.
\newblock Subextensive scaling in the athermal, quasistatic limit of amorphous
  matter in plastic shear flow.
\newblock {\em Phys. Rev. Lett.}, 93:016001, 2004.

\bibitem{maloney2006}
C.~E. Maloney and A.~Lema\^{i}tre.
\newblock Amorphous systems in athermal, quasistatic shear.
\newblock {\em Phys. Rev. E}, 74:016118, Jul 2006.

\bibitem{elmatad2010}
Y.~S. Elmatad, R.~L. Jack, D.~Chandler, and J.~P. Garrahan.
\newblock Finite-temperature critical point of a glass transition.
\newblock {\em Proc. Nat. Acad. Sci.}, 107:12793--12798, 2010.

\bibitem{pinney2016}
R.~Pinney, T.~B. Liverpool, and C.~P. Royall.
\newblock Structure in sheared supercooled liquids: Dynamical rearrangements of
  an effective system of icosahedra.
\newblock {\em J. Chem. Phys.}, 145:234501, 2016.

\end{thebibliography}
\end{document}